\documentclass[aps,prd,10pt,nofootinbib,
twocolumn, unsortedaddress]{revtex4-1}

\usepackage{epsfig}
\usepackage{bm}
\usepackage{latexsym}
\usepackage{natbib}
\usepackage{url}
\usepackage{dcolumn}
\usepackage{color}
\usepackage[usenames,dvipsnames]{xcolor}
\usepackage{amsfonts,amssymb,amsmath}
\usepackage{graphicx,epsfig}
\usepackage{psfrag}
\usepackage{subfigure}
\usepackage{comment}
\usepackage[utf8]{inputenc}
\usepackage{amsmath}
\usepackage{hyperref}
\hypersetup{colorlinks=true}

\def\be{\begin{equation}}
\def\ee{\end{equation}}
\def\gev{{\rm \,Ge\kern-0.125em V}}

\begin{document}
 
\title{Lowering tensor-to-scalar ratio by pre-inflationary dynamics}

\author{Suratna Das}%
 \email{suratna@iitk.ac.in}
\affiliation{Department of Physics, Indian Institute of Technology, Kanpur 208016, India}

\date{\today}

\begin{abstract}
Pre-inflationary dynamics is known to have left observable imprints via the varied dynamics of the inflaton and the graviton fields during those epochs. It is highlighted here that other particle degrees of freedom, too, are capable of leaving their imprints on primordial observables if they transfer their entropies upon annihilation after becoming non-relativistic during a pre-inflationary epoch. Such annihilations of other particles would result in lowering of the tensor-to-scalar ratio at the pivot scale (and at length scales larger to that) which would help bring those potentials back in accordance with current observations which are otherwise disfavoured for yielding large tensor-to-scalar ratios at present. 
\end{abstract}

\maketitle

\section{Introduction}

The high precision data of the Cosmic Microwave Background (CMB) temperature fluctuations probe only the 8--10 $e$-foldings of the nearly 60 $e$-foldings of accelerated expansion during cosmic inflaiton \cite{Akrami:2018odb}. Though, over the years the smaller scales of the primordial perturbations associated with later $e$-foldings have been measured with better precision, scales larger than what have already been observed are simply beyond any measurement. Despite such observational obstacle, study of pre-inflationary dynamics has remained a subject of interest in the literature. The theoretical motivation of studying such dynamics is lying in the fact that it is rather difficult to sustain inflation for a longer period which often calls for certain fine-tunings \cite{Hawking:1987bi, Freivogel:2005vv, Gibbons:2006pa}. Hence, the preference for `just-enough' inflation, providing a window for observational effects of pre-inflationary dynamics on the CMB.

Effects of quantum gravity theories, such as loop-quantum gravity \cite{Agullo:2013ai} and string theory \cite{Dudas:2012vv, Gruppuso:2015xqa}, playing a role during pre-inflationary era have been studied in the literature. However, effects of pre-inflationary radiation era has been studied most extensively in the context of pre-inflationary dynamics \cite{Powell:2006yg, Marozzi:2011da, Hirai:2002vm, Hirai:2003dh, Hirai:2004kh, Hirai:2005tn, Hirai:2007ne, Wang:2007ws}. Effects of more general equation of states have also been explored \cite{Cicoli:2014bja, Cai:2015nya}. The common effect of such pre-inflationary dynamics on the temperature anisotropy spectrum is a power loss at largest  angular scales, which is in accord with the observations since COBE \cite{Hinshaw:1996ut}, then WMAP \cite{Spergel:2003cb} and now PLANCK \cite{Ade:2013uln}. This can also be accounted for as an observational motivation of studying such pre-inflationary dynamics. Abruptly matching of inflaton wavefunctions at the boundary of pre-inflationary era and the era of slow-roll dynamics often gives rise to `ringing effects' in the lower multipoles of the temperature anisotropy spectrum \cite{Powell:2006yg, Marozzi:2011da, Hirai:2002vm, Hirai:2003dh, Hirai:2004kh, Hirai:2005tn, Hirai:2007ne}. A gain in power at the largest angular scales can occur if the inflaton field thermalises before decoupling from the cosmic soup during a pre-inflationary radiation era \cite{Bhattacharya:2005wn}. As the current data do not suggest such an enhancement, it puts a bound on the comoving temperature of the inflaton field. At large, these are the noted effects of pre-inflationary dynamics on the CMB temperature anisotropy spectrum. 

Pre-inflationary dynamics can also leave its imprints on the primordial tensor spectrum. The non-Bunch Davies vacuum during pre-inflationary era resulted from the loop-quantum cosmology can lead to a modified consistency relation (relation between the tensor-to-scalar ratio and the tensor spectral tilt)  \cite{Agullo:2013ai}. Stringy effects during pre-inflationary era can enhance the power in the lower multipoles of the tensor spectrum \cite{Dudas:2012vv}. A thermalised graviton from a pre-inflationary radiation era enhances the power of the lowest multipole in the tensor spectrum \cite{Bhattacharya:2006dm, Zhao:2009pt}. Effects of a generic equation of state during pre-inflationary epoch on the primordial gravitational waves have been analyzed in \cite{Wang:2016wio}.

However, all these observable imprints result from the dynamics of the inflaton and the graviton fields during a pre-inflationary era. None of these imprints can probe the presence of other particle degrees of freedom which might have been present in the pre-inflationary cosmic plasma. Hence, it is natural to ask, what observable imprints other particle degrees of freedom could have possibly left on the CMB observations. More importantly, whether or not other particle degrees of freedom are at all capable of leaving any imprint on CMB observables. We will highlight in this {\it letter} that there is one such scenario where annihilation of other particle degrees of freedom during pre-inflationary epoch could, in principle, leave their imprints by lowering the tensor-to-scalar ratio during cosmic inflation. Such an imprint is yielded when the gravitons and the inflaton are both thermalised during a pre-inflationary era and decouple from the cosmic plasma at different epochs.

\section{Brief review of the dynamics of pre-inflationary radiation era}

To illustrate this, we first briefly review the dynamics of thermalised inflaton and gravitons in a pre-inflationary era. 
It was first noted in \cite{Bhattacharya:2005wn} that if inflation is preceded by a radiation era then the inflaton field might have been in thermal equilibrium with the pre-inflationary cosmic plasma due to its gravitational and other couplings. Even if the inflaton field decouples from this plasma prior to inflation it would retain a thermal distribution which would be reflected in its power spectrum. 

If the inflaton perturbations $\delta\phi({\mathbf x},\tau)$ ($\tau$ being the conformal time) are expanded in the Fourier space 
then due to the thermal distribution the modes will have an occupation number:
\begin{eqnarray}
\langle a^\dagger_{\mathbf k}a_{\mathbf k'}\rangle=\frac{1}{e^{E_k/T_{d}^{\rm inf}}-1}\delta^3(\mathbf{k}-{\mathbf k}'),
\end{eqnarray}
where ${\mathbf k}$ is the comoving wavenumber, $T_{d}^{\rm inf}$ is the (physical) decoupling temperature of the inflaton field and $E_k$ is the corresponding energy of the $k$ mode. As the temperature of a decoupled species scales as $ T\propto 1/a$ and $E_k=k/a_d$ ($a_d$ being the scale factor at decoupling), one has $E_k/T_{d}^{\rm inf}=k/(a_d T_{d}^{\rm inf})=k/(a_{\rm ini} T^{\rm inf}_{\rm ini})=k/{\mathcal T}^{\rm inf}$, where $a_{\rm ini}$ and $T_{\rm ini}^{\rm inf}$ are the scale factor and physical temperature at the onset of inflation and ${\mathcal T}^{\rm inf}$ is the comoving temperature of the inflaton perturbations.\footnote{Note that in \cite{Bhattacharya:2005wn} $\mathcal T$ and $T$ have been used to represent the physical and comoving temperatures respectively, opposite to the notations adopted here.}

On the other hand, in the vanilla model of single field inflation (where the scenario is not preceded by a radiation era) the occupation number of the inflaton modes would read as $\langle a^\dagger_{\mathbf k}a_{\mathbf k'}\rangle=0$ in vacuum. Because of this very difference between these two scenarios, the scalar amplitude in this case of thermalised inflaton would get modified by a multiplicative $\coth$ factor as \cite{Bhattacharya:2005wn}
\begin{eqnarray}
A_s&=&\frac{1}{24\pi^2\epsilon}\frac{V}{M_{\rm Pl}^4}\coth\left(\frac{k}{2{\mathcal T}^{\rm inf}}\right),
\label{scalar-amp}
\end{eqnarray}
enhancing the power at the largest scales or for the smallest wavenumbers $k$.\footnote{It was pointed out in \cite{Das:2014ffa, Das:2015ywa} that if the matching of the scalar modefunctions at the boundary of transition from pre-inflationary era to inflationary epoch is taken into account along with the thermal distributions of inflaton modes then the lowering effect of the modefunction matching cancels out the enhancing of the power at the largest scales due to thermal distribution, leaving negligible effect on the CMB temperature anisotropy spectrum. However, this is of no interest in the present analysis.}
It is to note that, as the modification through the $\coth$ factor depends on comoving wavenumber $k$ and comoving temperature $\mathcal{T}$ of the modes alone, which do not depend on the number of $e$-foldings, such an effect {\it will not} get diluted away in scenarios beyond `just enough' inflation. Hence, such a signature will sustain even if inflation lasts longer. The signature of such thermalised inflaton on primordial non-Gaussianities has been studied in \cite{Das:2009sg}.

Similarly, gravitons, which interact only gravitationally with the cosmic plasma, might have been decoupled from the pre-inflationary radiation bath near the Planck era and would have retained a thermal distribution, same as the inflaton field. The tensor perturbations during inflation would then be generated by stimulated emissions into this pre-existing thermal background of gravitons. Effects of such thermal gravitons on the primordial tensor perturbations (the $BB$ spectrum of CMB) has been studied in  \cite{Bhattacharya:2006dm}. As the two polarization states, $h^+$ and $h^\times$, of the primordial gravitational waves evolve independently, and each of them behaves like a massless scalar field, expanding each of these two polarizations ($i\equiv +,\times$) in Fourier space as 
\begin{eqnarray}
h^{(i)}({\mathbf x},\tau)&=&\frac{\sqrt{2}}{a(\tau)M_{\rm Pl}}\times \nonumber\\
&&\int \frac{d^3k}{(2\pi)^{\frac32}}\left[b_{\mathbf k}^{(i)}f_k(\tau)+b^{(i)\dagger}_{-\mathbf k}f^*_k(\tau)\right]e^{i{\mathbf k}\cdot{\mathbf x}},
\end{eqnarray}
it is straightforward to note that each of these polarization modes would have an occupation number 
\begin{eqnarray}
\langle b^\dagger_{\mathbf k}b_{\mathbf k'}\rangle=\frac{1}{e^{k/{\mathcal T}^{\rm grav}}-1}\delta^3(\mathbf{k}-{\mathbf k}'),
\end{eqnarray}
where ${\mathcal T}^{\rm grav}$ is the comoving temperature of the tensor modes, 
resulting in a similar multiplicative $\coth$ term in the primordial tensor amplitude: 
\begin{eqnarray}
A_T&=&\frac{2}{3\pi^2}\frac{V}{M_{\rm Pl}^4}\coth\left(\frac{k}{2{\mathcal T}^{\rm grav}}\right),
\label{tensor-amp}
\end{eqnarray}
as has been shown in \cite{Bhattacharya:2006dm}. As in the case of scalars, signatures of such a modification to the tensor power spectrum, too, would sustain even when inflation lasts longer.  

\section{Lowering of the tensor-to-scalar ratio} 

If both the gravitons and the inflaton fields interact with the plasma through gravitational interactions alone and thus decouple at the Planck scale, then in such a case, the decoupling temperatures of both the graviton and inflaton fields would be the same. Even though they evolve as two separate species thereafter, both their temperatures would scale as $1/a$,\footnote{It is to note that here it has been assumed that the pre-inflationary epoch can be described by as a standard FRW radiation dominated era, which has been also considered in the literature previously \cite{Powell:2006yg, Marozzi:2011da, Hirai:2002vm, Hirai:2003dh, Hirai:2004kh, Hirai:2005tn, Hirai:2007ne, Wang:2007ws, Das:2014ffa, Das:2015ywa}. The consequences of such an assumption has been discussed at length in \cite{Das:2014ffa}.} yielding them to have the same temperature at the onset of inflation. But, if the inflaton field has interactions beyond the mere gravitational ones with the pre-inflationary cosmic plasma, then it is logical to assume that the inflaton field would decouple at a much later stage than the gravitons. In such a case, the decoupling temperatures of the inflaton and the graviton fields would be different. However, if no other species of the cosmic plasma annihilate upon becoming non-relativistic and transfer their entropy to the bath in between the decouplings of the gravitons and the inflaton, then the temperature of the decoupled gravitons and inflaton would be the same at the onset of inflation. This is due to the conservation of entropy because of which the temperature of the cosmic plasma scales as $T\propto g_{*S}^{-1/3}a^{-1}$, where $g_{*S}$ is the relativistic degrees of freedom associated with entropy density. If none of the other species annihilate then $g_{*S}$ would remain invariant throughout the evolution before inflaton decouples, hence the temperature of the cosmic plasma as well as that of the inflaton in local thermal equilibrium would scale as $T\propto a^{-1}$, same as that of the decoupled gravitons. 

In both these scenarios when the gravitons and the inflaton retain the same temperature at the onset of inflation, one has at the pivot scale $T^{\rm inf}_*=T^{\rm grav}_*$ or ${\mathcal T}^{\rm inf}={\mathcal T}^{\rm grav}$. In such a case, the $\coth$ factors contributing to the scalar and tensor amplitudes (as given in Eq.~(\ref{scalar-amp}) and Eq.~(\ref{tensor-amp}) respectively) would render to be the same, yielding a tensor-to-scalar ratio at the pivot scale:
\begin{eqnarray}
r_*=16\epsilon,
\end{eqnarray}
and leaving no imprint of pre-inflationary dynamics on this CMB observable. 

However, the gravitons and the inflaton would evolve with different temperatures if some of the species annihilate and transfer their entropies to the cosmic plasma in between the decouplings of the gravitons and the inflaton. In such a case, the tensor-to-scalar ratio would then turn out to be 
\begin{eqnarray}
r_*=16\epsilon\left[\coth\left(\frac{k_*}{2{\mathcal T}^{\rm grav}}\right)/\coth\left(\frac{k_*}{2{\mathcal T}^{\rm inf}}\right)\right],
\label{t-to-s}
\end{eqnarray}
where $k_*$ is the pivot scale. Such an effect would also not get diluted away if inflation lasts longer than the bare minimum amount of $e$-folds. 

In \cite{Bhattacharya:2005wn}, the pivot scale has been chosen at $k_*=0.05$ Mpc$^{-1}$, and using the WMAP data\footnote{It is to note that the $\coth$ factor affects only the largest scales or the lowest multipoles and leave the smaller scales unaffected as $\coth(x)\sim1$, when $x\gg1$. As the error bars in the PLANCK data for the lowest multipoles receives not much improvement from the previous WMAP data, we do not expect any improvement in the bound on the comoving temperature obtained in  \cite{Bhattacharya:2005wn}.} a bound on the comoving temperature of the inflaton field has been derived as 
\begin{eqnarray}
{\mathcal T}^{\rm inf}< 1.0\times 10^{-3} \, {\rm Mpc}^{-1}.
\end{eqnarray}
The tensor-to-scalar ratio, on the other hand,  is generally quoted at a pivot scale $k_*=0.002$ Mpc$^{-1}$ \cite{Akrami:2018odb}, and a bound on graviton comoving temperature has been considered in \cite{Bhattacharya:2006dm} as
\begin{eqnarray}
{\mathcal T}^{\rm grav}< 1.0\times 10^{-3} \, {\rm Mpc}^{-1},
\end{eqnarray}
same as that of the inflaton modes. As the BB anisotropy spectrum is yet to be observed, obtaining a direct upper bound on the comoving temperature of the tensor modes is beyond the scope. However, such a consideration is well justified as in general the decoupled gravitons and the inflatons would have similar comoving temperature (at least of the same order) at the onset of inflation. 
We will consider, in our later derivations, that the bound on the inflaton comoving temperature will remain the same if we shift the pivot scale from 0.05 Mpc$^{-1}$ to 0.002 Mpc$^{-1}$, as the comoving temperature of the inflaton field would not vary with the wavenumber $k$. 
It is to note that $\coth(x)$ can be expanded as $\coth(x)\sim1/x+\cdots$
when $0<|x|<\pi$. Now, taking $k_*=0.002$ Mpc$^{-1}$, we have $k_*/2{\mathcal T}^{\rm inf, grav}=1$ if we take ${\mathcal T}^{\rm inf, grav}=0.001$ Mpc$^{-1}$. This would allow us to expand the $\coth$'s in Eq.~(\ref{t-to-s}) at the pivot scale $k=0.002$ Mpc$^{-1}$  to have
\begin{eqnarray}
r_{0.002}\approx 16\epsilon  \left( \frac{{\mathcal T}^{\rm grav}}{{\mathcal T}^{\rm inf}}\right). 
\label{t-to-s1}
\end{eqnarray}
As ${\mathcal T}^{\rm grav}<{\mathcal T}^{\rm inf}$ due to the annihilation of the in-between particle degrees of freedom, the tensor-to-scalar ratio would be lowered than the usual cold-inflation case. We will now see how ${\mathcal T}^{\rm grav}$ and ${\mathcal T}^{\rm inf}$ are related in a simple scenario when the species annihilate in between around the same time. 

Let us assume that at physical temperature $T^{\rm grav}_d$ the gravitons decoupled from the cosmic soup. Some time later at $t=t_1$, some species become non-relativistic, annihilate in equilibrium and transfer their entropy to the cosmic plasma instantaneously. We denote the (physical) temperature of the plasma before and after the annihilation of these particles as $T_b$ and $T_a$, respectively. Conservation of entropy would imply then 
\begin{eqnarray}
\frac{T_b}{T_a}=\left(\frac{g_{*a}}{g_{*b}}\right)^{1/3},
\end{eqnarray}
where $g_{*b}$ and $g_{*a}$ are the relativistic degrees of freedom before and after the annihilation of the particles, respectively. 

If no other species annihilates before the inflaton decouples at the temperature $T^{\rm inf}_d$, then we have for the temperature of the graviton at the onset of inflation as: 
\begin{eqnarray}
T^{\rm grav}_{\rm ini}=\frac{a^{\rm grav}_d}{a_{\rm ini}}T^{\rm grav}_d,
\end{eqnarray}
where $a^{\rm grav}_d$ and $a_{\rm ini}$ are scale factors at the decoupling of the graviton and at the onset of inflation. 

On the other hand, for the inflaton field we have 
\begin{eqnarray}
T^{\rm inf}_{\rm ini}=\frac{a_a}{a_{\rm inf}}T_a= \frac{a_a}{a_{\rm inf}}T_b\left(\frac{g_{*b}}{g_{*a}}\right)^{1/3},
\end{eqnarray}
where $a_a$ is the scale factor after the annihilation of the particles. We also see that $a_d^{\rm grav}T^{\rm grav}_d=a_bT_b$ ($a_b$ being the scale factor before the annihilation of the particles), including which in the above equation we have
\begin{eqnarray}
T^{\rm inf}_{\rm ini}=\frac{a_a}{a_b}\frac{a^{\rm grav}_d}{a_{\rm ini}}T^{\rm grav}_d\left(\frac{g_{*b}}{g_{*a}}\right)^{1/3}.
\end{eqnarray}
As the particles annihilates and transfer their entropy instantaneously at $t_1$ we have $a_a=a_b$. Thus the above equation yields 
\begin{eqnarray}
T^{\rm inf}_{\rm ini}=\left(\frac{g_{*b}}{g_{*a}}\right)^{1/3}T^{\rm grav}_{\rm ini},
\end{eqnarray}
which also implies that 
\begin{eqnarray}
{\mathcal T}^{\rm inf}=\left(\frac{g_{*b}}{g_{*a}}\right)^{1/3}\mathcal{T}^{\rm grav}.
\end{eqnarray}
Inserting this into Eq.~(\ref{t-to-s1}) one gets 
\begin{eqnarray}
r_{0.002}\approx 16\epsilon \left(\frac{g_{*a}}{g_{*b}}\right)^{1/3}.
\end{eqnarray}
As $g_{*a}<g_{*b}$, the tensor-to-scalar ratio would be lowered than in the generic cold inflation case.


This phenomenon of lowering of tensor-to-scalar ratio due to pre-inflationary dynamics can be exploited in those cases where the potentials yield way too large tensor-to-scalar ratios to be in accordance with the observations, a classic example of which is the well-known monomial potentials appearing in chaotic inflation \cite{Akrami:2018odb}. 
Let us consider the case of quartic potential which is now disfavoured by the current data in the case of generic cold inflation \cite{Akrami:2018odb}. The reason for this is as follows. 
For the quartic potential, $V(\phi)=\lambda\phi^4$, we get $\epsilon=8M_{\rm pl}^2/\phi^2$ and $N=\phi^2/8M_{\rm pl}^2$, $N$ being the total number of efoldings. Thus we have 
\begin{eqnarray}
r_*\sim 16\epsilon=16/N\sim 0.267,
\end{eqnarray}
for $N=60$. The current bound on the tensor-to-scalar ratio is $r_{0.002}<0.064$ \cite{Akrami:2018odb}, which clearly disfavours the quartic inflaton potential. But in the above discussed scenario, the quartic potential can be made in tune with the present data if 
\begin{eqnarray}
\left(\frac{g_{*a}}{g_{*b}}\right)^{1/3}<0.064\times \frac{N}{16}\sim 0.24,
\end{eqnarray}
for $N=60$. If we assume nearly 20 species of pair of fermions (spin $1/2$ particles) annihilates transferring their entropy to the the cosmic plasma leaving the inflaton alone to sustain then we get $g_{*b}=1+(7/8)\times 4\times 20=71$, and $g_{*a}=1$, leading to the above factor. Similarly, for quadratic potential, one has $\epsilon\sim 1/2N$, which yields $r_{0.002}\sim0.133$. Thus, quadratic potentials could be made in accordance with observations if 
\begin{eqnarray}
\left(\frac{g_{*a}}{g_{*b}}\right)^{1/3}<0.064\times \frac{N}{8}\sim 0.48,
\end{eqnarray}
for $N=60$. In this case, only three such fermionic species, like in the case of quartic potential, are required to annihilate and transfer their entropies to the plasma (yielding $g_{*b}=23/2$ and $g_{*a}=1$) in order to achieve the above bound. 

\section{Conclusion}
{\it To summarize}, it is shown that the other particle degrees of freedom, apart from the inflaton and the gravitons, present in the pre-inflationary cosmic plasma are capable of leaving observable imprints on the CMB observables. If these particles transfer their entropy to the pre-inflationary cosmic plasma upon annihilation after becoming non-relativistic, then they would result in a temperature difference between the gravitons and the inflaton field when the inflaton decouples from the plasma afterwards. This temperature difference would result in lowering of the tensor-to-scalar ratio measured at a pivot scale at 0.002 Mpc$^{-1}$. Such an effect would help bring the potentials like quadratic and quartic chaotic potentials, which are otherwise ruled out for yielding way too large tensor-to-scalar ratios, within the range of the current observations. Moreover, such an effect can provide glimpses of the rich dynamics of a pre-inflationary thermal era, which has not been reported before to our knowledge.


 {\it Acknowledgements:} The work of S.D. is supported by Department of Science and Technology, 
 Government of India under the Grant Agreement number IFA13-PH-77 (INSPIRE Faculty Award). The author is grateful to Subhendra Mohanty, Kaushik Bhattacharya and Raghavan Rangarajan for useful comments on the draft. 
 
\label{Bibliography}
\bibliography{pre-inf}

\end{document}